# Time-slicing spiral galaxies with SDSS-IV MaNGA

Thomas Peterken,[1]⋆ Amelia Fraser-McKelvie,[1] Alfonso Aragón-Salamanca,[1] Michael Merrifield,[1] Katarina Kraljic,[2] Johan H. Knapen,[3,4,5] Rogério Riffel,[6,7] Joel Brownstein[8] and Niv Drory[9]

[1]*School of Physics and Astronomy, University of Nottingham, University Park, Nottingham NG7 2RD, UK*
[2]*Institute for Astronomy, University of Edinburgh, Royal Observatory, Blackford Hill, Edinburgh EH9 3HJ, UK*
[3]*Instituto de Astrofísica de Canarias (IAC), La Laguna, E-38205, Spain*
[4]*Departamento de Astrofísica, Universidad de La Laguna (ULL), E-38200, La Laguna, Spain*
[5]*Astrophysics Research Institute, Liverpool John Moores University, IC2, Liverpool Science Park, 146 Brownlow Hill, Liverpool, L3 5RF, UK*
[6]*Departamento de Astronomia, IF, Universidade Federal do Rio Grande do Sul, CP 15051, 91501-970, Porto Alegre, RS, Brazil*
[7]*Laboratório Interinstitucional de e-Astronomia - LIneA, Rua General José Cristino 77, Rio de Janeiro, RJ - 20921-400, Brazil*
[8]*Department of Physics and Astronomy, University of Utah, 115 S. 1400 E., Salt Lake City, UT 84112, USA*
[9]*McDonald Observatory, The University of Texas at Austin, 2515 Speedway Stop C1402, Austin, TX 78712, USA*

**ABSTRACT**

Spectra of galaxies contain a wealth of information about the stellar populations from which they are made. With integral-field unit (IFU) surveys, such data can be used to map out stellar population properties across the face of a galaxy, allowing one to go beyond simple radial profiles and study details of non-axisymmetric structure. To-date, however, such studies have been limited by the quality of available data and the power of spectral analysis tools. We now take the next step and study the barred spiral galaxy MCG+07-28-064 from observations obtained as part of the SDSS-IV MaNGA project. We find that we can decompose this galaxy into "time slices," which reveal the varying contributions that stars of differing ages make to its bar and spiral structure, offering new insight into the evolution of these features. We find evidence for the ongoing growth of the bar, including the most recent star formation on its leading edge, and for the underlying density wave responsible for spiral structure. This pilot study indicates that there is a wealth of untapped information on the spatial distribution of SFHs available in the current generation of IFU galaxy surveys.

## 1 INTRODUCTION

By analyzing the full spectrum of an unresolved stellar population, it is possible to extract detailed information about the population's star formation history (SFH) by measuring the fraction of stars of different ages present. Early works by Panter et al. (2003), Ocvirk et al. (2006), Mathis et al. (2006), Tojeiro et al. (2007), Koleva et al. (2009) and others showed that spectral analysis of large low-redshift spectroscopic survey data can reveal information about the evolution of star formation in the Universe back to redshifts of $z \approx 2$.

With the advent of large integral-field spectroscopic surveys such as the Calar Alto Legacy Integral Field Area survey (CALIFA; Sánchez et al. 2012), the Sydney-AAO Multi-object Integral field spectrograph (SAMI; Croom et al. 2012), and Mapping Nearby Galaxies at Apache Point Observatory (MaNGA; Bundy et al. 2015), the populations of different regions of galaxies can be characterised. González Delgado et al. (2015) showed that the SFHs of galaxies vary with radius and with galaxy morphology. They found evidence for inside-out growth of galaxies, an effect which Ibarra-Medel et al. (2016) found to be strongest in star-forming spiral galaxies. Backing up these findings, Rowlands et al. (2018) also showed that the fractions of stars of different ages residing in different regions of a galaxy are connected to the local gas conditions in a galaxy. Measuring the spatially-resolved SFHs with MaNGA can also help in un-derstanding the known evolution of galaxy populations over cosmic time (Sánchez et al. 2019) as well as the effects of AGN activity (Mallmann et al. 2018).

⋆ E-mail: Thomas.Peterken@nottingham.ac.uk

Statistical studies using large samples are useful to help us understand the general evolution and growth of galaxies, but by only measuring variation in the SFHs with galactic radius some subtle details are being "washed out". Most low-redshift disc galaxies host non-axisymmetric structures such as spiral arms and bars (Knapen et al. 2000; Eskridge et al. 2000; Masters et al. 2011) which are likely to affect different stellar populations in different ways. It is known that azimuthal variation in the current star formation rate can arise as a result of the presence of bars (Martin & Friedli 1997; Verley et al. 2007) and spiral arms (Gonzalez & Graham 1996; Puerari & Dottori 1997; Dametto et al. 2019). Additionally, bars are associated with stable orbits (Contopoulos & Papayannopoulos 1980; Athanassoula 1992) and hence flattened stellar population gradients (Sánchez-Blázquez et al. 2014, Fraser-McKelvie et al. in prep.), while spiral arms are associated primarily with young stars (Roberts 1969).

To gain a more complete picture of the spatially resolved SFHs of disc galaxies beyond the radial variations of previous large-scale works, we should generate maps of stars of different ages and test how these correlate with galaxy components including non-axisymmetric structures. Cid Fernandes et al. (2013, 2014) showed that it is possible to observe the distinct structures of stellar populations of different ages using spectral synthesis of CALIFA datacubes. In this work, we aim to test how similar methods could be reliably applied to a MaNGA datacube in preparation for larger studies, and to also investigate whether non-axisymmetric structures vary significantly between different stellar populations. To this end, we perform a pilot study of full-spectral fitting on a datacube to find the fraction of light and mass in stars of different ages and metallicities. We take care to preserve the full spatial information available in MaNGA, in order to build maps of stars of different ages.

## 2 DATA

### 2.1 The MaNGA Survey

The MaNGA survey (Bundy et al. 2015) is part of the fourth generation of the Sloan Digital Sky Survey (SDSS-IV; Blanton et al. 2017), using the BOSS spectrograph (Smee et al. 2013) on the 2.5m Sloan telescope based at Apache Point Observatory (Gunn et al. 2006). By the project's completion in 2020, integral field spectroscopic data with a spatial resolution of 2.5 arcsec will have been obtained for more than 10,000 nearby ($0.01 < z < 0.15$) galaxies (Yan et al. 2016b), which will have a roughly flat log($M_\star$) distribution of galaxies more massive than $10^9$ $M_\odot$ (Wake et al. 2017). The spectra cover a very broad wavelength range of 3600-10300 Å, with a spectral resolution of $R \approx 2000$. The size of integral field unit (IFU) used for each galaxy is chosen so that each galaxy in the primary sample has coverage out to 1.5 $R_e$ (Law et al. 2015), with a choice of five different sizes varying from the 19-fibre IFUs with a diameter 12 arcsec, to the 127-fibre IFUs of 32 arcsec diameter (Drory et al. 2015).

The raw datacubes are calibrated (Yan et al. 2016a) and reduced by a Data Reduction Pipeline (DRP; Law et al. 2016), and some analysis products such as stellar kinematics, emission-line properties, and spectral indices are provided by a Data Analysis Pipeline (DAP; Westfall et al. 2019).

### 2.2 Galaxy MCG+07-28-064

For the purposes of this exploratory Letter, we have selected MCG+07-28-064 (MaNGA plate-IFU 8332-12701) as a example galaxy to test our population analysis techniques. It serves as an ideal test-case galaxy for this purpose, since it shows clear spiral and bar structure. According to the NASA-Sloan Atlas (NSA), it has a redshift of $z = 0.028$, and a mass of $10^{9.78}$ $M_\odot$ (Blanton et al. 2011).

As well as being in the MaNGA sample, MCG+07-28-064 is part of the initial sample of galaxies for Galaxy Zoo:3D programme (Masters et al. in prep.). In that project, citizen scientist volunteers are asked to draw bar, spiral arm, and other regions on images of MaNGA spiral galaxies. The collation of their results provides an independent assessment of the locations of bar and spiral arm features, which we can use in the subsequent analysis.

## 3 SPECTRAL FITTING

A number of spectral fitting tools are publicly available. In this work, we use STARLIGHT (Cid Fernandes et al. 2005), which has previously been successfully applied to analyse stellar populations of CALIFA and MaNGA datacubes (Cid Fernandes et al. 2013; Mallmann et al. 2018). For a given input spectrum, STARLIGHT utilises Monte Carlo techniques to find a best-fit combination of template spectra, also allowing for a variable-strength dust extinction.

We apply STARLIGHT to each spectrum in the MaNGA datacube of MCG+07-28-064 using STARLIGHT's "long fit" mode which prioritises robustness over speed. It is common to bin neighbouring spatial pixels ("spaxels") together when fitting to ensure a minumum signal-to-noise. However, to ensure that we are not blending any non-axisymmetric structures into the surrounding disc, we perform the fits on every available spaxel. This results in lower signal-to-noise in individual spaxels in the faint outskirts of the galaxy, but here we focus only on the brighter bar and spiral arm regions.

Since STARLIGHT does not fit emission lines, these must either be removed or masked before fitting. Although many authors (e.g. Cid Fernandes et al. 2013) choose to mask them, we make use of the MaNGA DAP (Westfall et al. 2019) emission-line measurements described by Belfiore et al. (2019) and subtract them from our spectra, allowing us to retain and fit all absorption line information which might otherwise be masked. The emission lines are fit using the MILES library of stellar – rather than stellar population – spectra to model the continuum, ensuring that there is no model assumption which might affect the STARLIGHT fits to the emission-subtracted spectra. The DAP's fitting and uncertainties associated with template choice are fully described by Belfiore et al. (2019) and Westfall et al. (2019).

The template spectra we use in fitting are synthetic single stellar populations (SSP) from E-MILES (Vazdekis et al. 2016). We use a set of 78 SSPs, with 14 ages (0.03, 0.05, 0.08, 0.1, 0.2, 0.3, 0.45, 0.7, 1, 1.75, 2.5, 4, 6.5, and 10 Gyr) and 6

metallicities ([M/H] = -1.79, -1.49, -0.96, -0.66, -0.25, +0.26), avoiding the six templates which fall outside the recommended safe ranges specified by Vazdekis et al. (2016). We have tested how our derived SFHs vary if we include a finer sampling in age and found that reducing the number of SSPs from 300 (the maximum allowed in STARLIGHT) to 78 resulted in entirely consistent results, but significantly reduced computing times. The templates assume a Kroupa (2001) IMF, Pietrinferni et al. (2004) isochrones, and a Milky Way [$\alpha$/Fe] ("baseFe").

We allow STARLIGHT to fit for extinction using a Calzetti (2001) curve, and use a wavelength range for the fit of 3300 to 8900 Å to make greatest use of all of the spectral information available in both MaNGA and E-MILES.

We have tested using a smaller set of SSPs with varying alpha-enhancement on a few individual spaxel spectra and found that although our resulting metallicity distributions may be affected by limiting to a single [$\alpha$/Fe], the age distributions appear to be unaffected. We have also tested adding a set of theoretical templates from González Delgado et al. (2005) covering the range of 1 to 50 Myr to the E-MILES set, and found that the SFHs at times greater than 50 Myr to be generally unaffected. Future work will incorporate these younger templates to probe the structure of populations at a greater range of stellar ages, but to avoid complications and additional uncertainties related to mixing two different population libraries, the results here are using only the E-MILES library.

For each spectrum in MCG+07-28-064, STARLIGHT returns the weight assigned to each template in its best fit at a reference wavelength of 4020 Å. By retaining all of these weights at each position in the MaNGA datacube, we produce a 4D output cube ($x, y,$ age, metallicity) from the 3D ($x, y, \lambda$) MaNGA datacube. Using this output, we can therefore sum the relevant template weights to find the fraction of stars at a specific location within a particular metallicity or age range. Here we use the STARLIGHT SSP weights to build maps of stars of each age, and can also find the mean metallicity of those stars across the face of the galaxy.

From the STARLIGHT fit, we find the total stellar mass within the NSA aperture to be $10^{9.94}$ $M_\odot$; entirely consistent with the NSA photometric value.

There are many uncertainties involved in full spectrum fitting (Conroy 2013; Cid Fernandes et al. 2014), particularly in the ability to break the age-metallicity degeneracy, although the broad wavelength range of MaNGA spectra helps here. As a result, many spectral fitting tools such as pPXF (Cappellari & Emsellem 2004) include regularisation to force the resulting SFHs to have a smoothed distribution. Since STARLIGHT does not include regularisation, we smooth the grid of SSP weights to produce maps of stars of different ages to ensure we are not over-interpreting large weights given to a few SSPs (Cid Fernandes et al. 2005; Cid Fernandes 2018).

It is also worth noting that the metallicity of young stellar populations is not easily determined, since these populations do not show strong absorption features. For populations younger than $\approx 0.1$ Gyr, STARLIGHT assigns weights to all metallicities approximately evenly, resulting in all populations at these ages to appear to have a mean metallicity of $\approx -0.75$.

**Figure 1.** *Top*: Animation showing the spatially resolved flux (colour-coded by the metallicity) of stars as a function of age, from 10 to 0.03 Gyr. *Middle*: Weighting function used. The STARLIGHT output is smoothed to 0.3 dex. Red points indicate the SSP ages used. *Bottom*: Colour map indicating the flux (in units of $10^{-14}$ erg s$^{-1}$ cm$^{-2}$ Å$^{-1}$ spaxel$^{-1}$) and metallicity (in units of log($Z/Z_\odot$)) of the stellar population. Dashed vertical lines indicate the SSP metallicities used. Requires Adobe Reader version $\geq$ 9 (and $\leq$ 9.4.1 on Linux) or similar. A high-resolution version is available online as supplementary material.

## 4 RESULTS

We visualise the spectral fitting on the datacube for MCG+07-28-064 by the animation shown in Fig. 1, showing the flux from stars as a function of age in different parts of the galaxy, coloured by those stars' mean metallicity. For each frame, we interpolate over the SSP weights using a Gaussian of width 0.3 dex in age to create a flux map of the stars of each age. We then find the mean metallicity of the stars of each ages at each point in the galaxy, which is indicated using the colour of each pixel. The animation is also available online as supplementary material.

In the animation, stars of varying ages exhibit varying radial and azimuthal structures. The presence of such age-dependent structures provides some confidence in performing spectral fitting in a spaxel-by-spaxel manner. To highlight the importance of retaining the full spatial information from MaNGA, we focus here on the variation of azimuthal structure in populations of different ages.

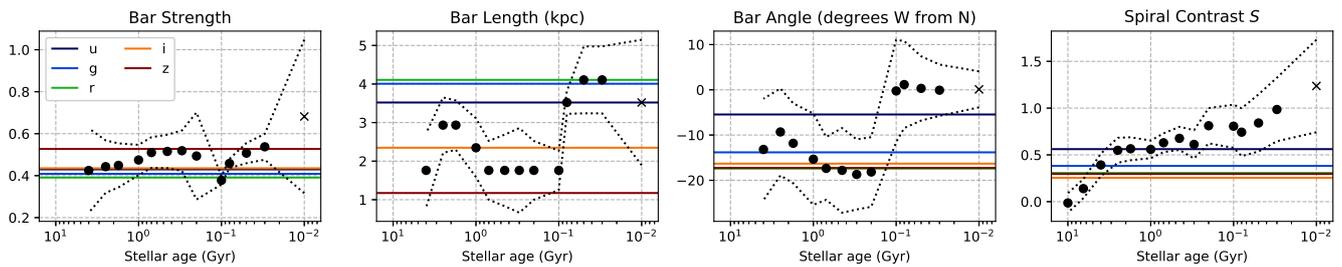

**Figure 2.** Bar and spiral parameters as a function of stellar age. Filled points are from the stellar population analysis, and the cross shows the results from Hα emission. Uncertainties are indicated by dotted lines. Coloured lines indicate the measurements obtained from the SDSS bands. Uncertainties in the bar parameters were estimated from the variations in measurements performed on each of the 4 halves (N,E,S,W) of the image independently, and uncertainties in the spiral contrast are estimated from the scatter in $f_s(r,t)$ and $f_d(r,t)$ in Eq. 1.

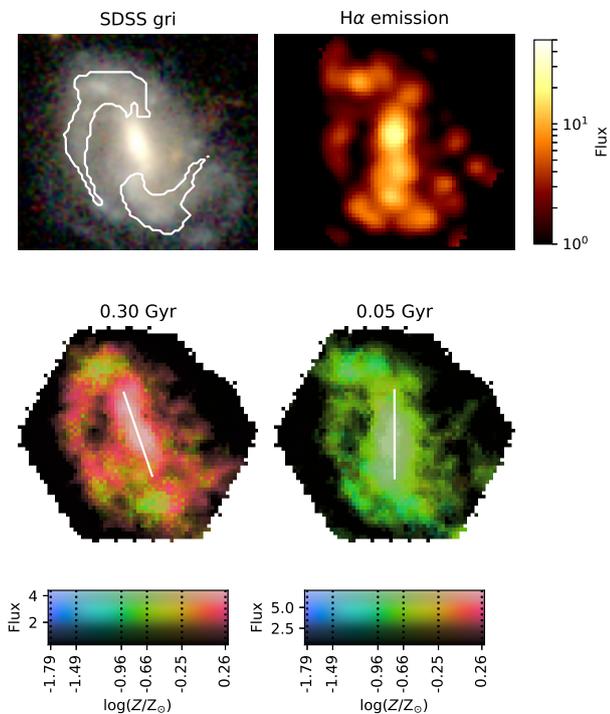

**Figure 3.** SDSS gri band image of galaxy MCG+07-28-064 (*top left*) showing outlines for the spiral mask from Galaxy Zoo:3D, Hα emission line flux map from the DAP (*top right*), and maps of the stellar populations centred on ages of 0.3 Gyr and 0.05 Gyr (*bottom left* and *bottom right* respectively), highlighting the observed change in bar angle between these stellar populations (indicated by the white lines). Both stellar maps are created by smoothing all SSP weights with a Gaussian of width 0.3 dex centred on the labelled age. In both frames for the stars, the flux (measured in $10^{-14}$ erg s$^{-1}$ cm$^{-2}$ Å$^{-1}$) and the mean metallicity (in $\log(Z/Z_\odot)$) of each spaxel is represented using the colour scheme shown underneath. The metallicities of the youngest stars are not trustworthy, as explained in Section 4.

To quantify these variations, we generated images of each stellar population time slice (smoothed to 0.3 dex in age), and measured parameters relating to the spiral arms and the bar. The bar structural parameters are measured using the Fast Fourier Transform bar analysis method of Kraljic et al. (2012). The presence of a bar is inferred by a strong $m = 2$ Fourier component. The length is inferred by measuring the phase of the second mode, $\Phi_2(r)$, within the bar region. A bar is present if $\Phi_2(r)$ is constant to within 5°, and the radii that this occurs for correspond to radii at which a bar is present. The length, strength, and angle of the bar are measured for age slices centred on each of the SSP ages used in the fitting.

To define the spiral arm contrast $S(t)$ for a given population age $t$, we first measure the spiral arm and disc flux profiles $f_s(r,t)$ and $f_d(r,t)$ as a function of radius $r$ at that age $t$. These are obtained using the Galaxy Zoo:3D spiral masks, where the spiral region is defined as that where at least 40% of users have marked as being part of the spiral arms, and the disk region as that where less than 20% of users marked as being spiral arms. These thresholds were chosen to ensure that a sufficient number of spaxels are included in both regions. Using these, we define a radially-dependent spiral contrast

$$S(r,t) = \frac{f_s(r,t) - f_d(r,t)}{f_d(r,t)}, \qquad (1)$$

and then take the median value of this quantity as the single-valued spiral arm contrast $S(t)$. Fig. 2 shows how these bar and spiral arm parameters change with stellar population age. The same parameters measured for the Hα emission map – which represents the very latest star formation in the galaxy – is consistent with those measured for the youngest populations. The SDSS u, g, r, i and z band images also show broad agreement with the results from the youngest to oldest stellar populations (as expected since the redder bands are more sensitive to older populations and vice-versa), but highlight the amount of extra information which can be obtained from this stellar population analysis.

The bar is detected in stellar populations of all stars younger than 4 Gyr, and *exhibits variation in its length at different stellar ages*. It is not clear that this variation necessarily reflects the historical bar growth, or whether it is an imprint of the original distribution of stars which became locked into the bar when it formed. However, since the bar instability is fundamentally a disk-related phenomenon, in either scenario it reflects the inside-out growth of the disk in this system. The other clear feature in the bar is the *discontinuity in its position angle at the youngest ages*, which we have highlighted in Fig.3. If the spiral arms are assumed to be trailing, this offset implies that stars are forming preferentially on the leading edge of the bar, as might be expected in this shock-forming region (Martin & Friedli 1997; Verley et al. 2007). The offset is only seen in stars younger than 0.1 Gyr because they have not had time to mix around their orbits, populating the whole bar potential. The agreement of the young stars' bar position angle with the independently-derived result from the H$\alpha$ map adds significant confidence in the stellar population analysis.

*A clear trend with age is also apparent in the spiral arm contrast*. In the oldest stars the arms are not detected, while at intermediate ages there appears to be about a 75% enhancement of stellar density in the arms, which rises in stars younger than $10^8$ years toward the H$\alpha$ contrast. Again, this picture is consistent with what might be expected in such a grand-design spiral galaxy, with the oldest stars forming a well-mixed population that is too dynamically hot to show the imprint of the arm, the intermediate-age stars making up the relatively modest density contrast representative of a spiral density wave, and the youngest stars showing the disproportionately-enhanced star formation that such a density wave can produce (Lin & Shu 1964; Roberts 1969).

The observed increases in bar length and angle in Fig. 2 at a stellar age of 2.5 Gyr are not real, and are a reflection of the level of systematic errors and degeneracies present in this analysis. Inspection of Fig. 1 at 2.5 Gyr indicates that these features are driven by an apparent population of low-metallicity stars of this age at the ends of the bar. The existence of such a trapped population would be fascinating, but it seems more likely that their presence arises from crosstalk between stellar population templates, using stars with these properties to fill in some systematic shortcoming in this complex fitting process.

## 5 DISCUSSION AND CONCLUSIONS

The above pilot analysis shows that there is a great deal of detailed information that can be derived on stellar populations from spectral fitting to IFU data. The coherence of the structures that emerge from this analysis, and their broad consistency with expectations in the context of galaxy evolution, indicate that our analysis allows us to extracting information from spectra that would otherwise not be available. It is important to note that this coherence is in no way imposed on the analysis – since the spaxels are fitted individually, if the process were just fitting noise then the lack of information would be reflected in largely featureless noisy images. Further reassurance is offered by the consistency between the spatial properties of the youngest stellar populations derived from the absorption-line spectra and the results obtained entirely independently from the H$\alpha$ emission-line image extracted from the same data cube.

We have highlighted some occasions where systematic errors and degeneracies when fitting so many parameters to each spectrum result in spurious features in the extracted time slices. More work needs to be done to understand and quantify the covariances that will exist in this kind of analysis, but this initial study shows that such effects are not likely to compromise the analysis of higher surface-brightness features if proper care is taken in the interpretation of the results.

Having established the viability of deriving time slices for the full two-dimensional structure of galaxies using MaNGA IFU data cubes in this pilot study, it would now seem timely to expand analysis of this type to the large sample of galaxies becoming available in this and other surveys. Detailed stellar population analysis that goes beyond simple radial gradients will allow us to understand the properties of azimuthal structures like bars and spiral arms in a way that has never been accessible before.

## ACKNOWLEDGEMENTS

We thank the anonymous referee for the constructive comments and suggestions to improve this manuscript.

Funding for the Sloan Digital Sky Survey IV has been provided by the Alfred P. Sloan Foundation, the U.S. Department of Energy Office of Science, and the Participating Institutions. SDSS-IV acknowledges support and resources from the Center for High-Performance Computing at the University of Utah. The SDSS web site is www.sdss.org.

SDSS-IV is managed by the Astrophysical Research Consortium for the Participating Institutions of the SDSS Collaboration including the Brazilian Participation Group, the Carnegie Institution for Science, Carnegie Mellon University, the Chilean Participation Group, the French Participation Group, Harvard-Smithsonian Center for Astrophysics, Instituto de Astrofísica de Canarias, The Johns Hopkins University, Kavli Institute for the Physics and Mathematics of the Universe (IPMU) / University of Tokyo, Lawrence Berkeley National Laboratory, Leibniz Institut für Astrophysik Potsdam (AIP), Max-Planck-Institut für Astronomie (MPIA Heidelberg), Max-Planck-Institut für Astrophysik (MPA Garching), Max-Planck-Institut für Extraterrestrische Physik (MPE), National Astronomical Observatories of China, New Mexico State University, New York University, University of Notre Dame, Observatário Nacional / MCTI, The Ohio State University, Pennsylvania State University, Shanghai Astronomical Observatory, United Kingdom Participation Group, Universidad Nacional Autónoma de México, University of Arizona, University of Colorado Boulder, University of Oxford, University of Portsmouth, University of Utah, University of Virginia, University of Washington, University of Wisconsin, Vanderbilt University, and Yale University.

This publication uses data generated via the Zooniverse.org platform, development of which is funded by generous support, including a Global Impact Award from Google, and by a grant from the Alfred P. Sloan Foundation. This publication has been made possible by the participation of almost 6000 volunteers in the Galaxy Zoo:3D project on Zooniverse.org.

We are grateful for access to the University of Nottingham High Performance Computing Facility.

J.H.K. acknowledges financial support from the Euro-pean Union's Horizon 2020 research and innovation programme under Marie Skłodowska-Curie grant agreement No 721463 to the SUNDIAL ITN network, from the Spanish Ministry of Economy and Competitiveness (MINECO) under grant number AYA2016-76219-P, from the Fundación BBVA under its 2017 programme of assistance to scientific research groups, for the project "Using machine-learning techniques to drag galaxies from the noise in deep imaging", and from the Leverhulme Trust through the award of a Visiting Professorship at LJMU.

RR thanks to CNPq, CAPES and FAPERGS for financial support to this project.